\documentclass[12pt]{article}

\usepackage{scicite}
\usepackage{amssymb, amsmath, verbatim, epsfig, url, color, caption, multicol, graphicx, float, caption, sectsty}
\subsectionfont{\normalfont\large\underline}

\usepackage{times}

\newcommand{\fig}[1]{{Fig. \ref{#1}}}
\newcommand{\figS}[1]{{Fig. S\ref{#1}}}

\DeclareCaptionType[name={Table S}]{supptable}
\DeclareCaptionType[name={Figure S}]{suppfig}

\newenvironment{sciabstract}{%
\begin{quote} \bf}
{\end{quote}}

\newcounter{lastnote}

\title{Sidewinding with minimal slip: Snake and robot ascent of sandy slopes}

\author
{Hamidreza Marvi$^{1,2}$, Chaohui Gong$^{4}$, Nick Gravish$^{2}$, Henry Astley$^{2}$, \\ \ Matthew Travers$^{4}$, Ross L. Hatton$^{6}$, Joseph R. Mendelson III$^{3,5}$, Howie Choset$^{4}$, \\
David L. Hu$^{1,2,3}$ \& Daniel I. Goldman$^{2\ast}$\\
\\
\normalsize{$^{1}$School of Mechanical Engineering, Georgia Institute of Technology,}\\
\normalsize{$^{2}$ School of Physics, Georgia Institute of Technology,}\\
\normalsize{$^{3}$ School of Biology, Georgia Institute of Technology,}\\
\normalsize{$^{4}$ Robotics Institute, Carnegie Mellon University, and}\\
\normalsize{$^{5}$ Department of Herpetology, Zoo Atlanta}\\
\normalsize{$^{6}$ School of Mechanical, Industrial, and Manufacturing Engineering, Oregon State University}\\
\\
\normalsize{$^\ast$To whom correspondence should be addressed;}\\ 
\normalsize{E-mail:  daniel.goldman@physics.gatech.edu}
} 

\date{}
\begin{document} 


\baselineskip24pt

\maketitle 

%% The abstract max is 125 words

\begin{sciabstract}
%Limbless organisms like snakes navigate most terrain. In particular, desert-dwelling sidewinder rattlesnakes operate effectively on granular materials such as sand; these substrates often stymie limbless robots. By discovering locomotor principles used in sidewinding and measuring yield forces in inclined granular media, we enable a field-tested limbless robot to ascend sandy slopes close to the angle of maximum slope stability. As incline angle increases, the snake increases the length of body in contact with the sand to remain below the decreasing granular yield stress. Implementing this control strategy in the robot avoids slipping and pitching failure modes, significantly improving performance. We propose that simultaneous study of biological and robotic locomotors coupled with investigations of the physics of complex materials is aligned with the original spirit of cybernetics, and can enable life-like locomotion in robots.

Limbless organisms like snakes can navigate nearly all terrain. In particular, desert-dwelling sidewinder rattlesnakes ({\em C. cerastes}) operate effectively on inclined granular media (like sand dunes) that induce failure in field-tested limbless robots through slipping and pitching. Our laboratory experiments reveal that as granular incline angle increases, sidewinder rattlesnakes increase the length of their body in contact with the sand. Implementing this strategy in a physical robot model of the snake enables the device to ascend sandy slopes close to the angle of maximum slope stability. Plate drag experiments demonstrate that granular yield stresses decrease with increasing incline angle. Together these three approaches demonstrate how sidewinding with contact-length control mitigates failure on granular media.

%The integration of biological and robotic studies will enable

%We propose that simultaneous study of biological and robotic locomotors coupled with investigations of the physics of complex materials 

\end{sciabstract}

% \section* for Review Articles tend to have displayed headings, for
% \paragraph* for Research Articles
% No space in \cite, only use commas

%%% INTRODUCTION
%Robots are poised to enter our everyday lives, transforming how we interact with the physical world. In the near future, multi-functional robots will advance scientific discovery across disciplines (e.g. archeology, biology, geology) \cite{marchant2012underwater}, aid first responders in search-and-rescue~\cite{murphybook} and accompany soldiers on the battlefield. Unlike aerial or aquatic systems (e.g., UAVs, ocean-monitoring submersible gliders, or salvage-and-exploration submersibles), 
The majority of terrestrial mobile robots are restricted to laboratory environments, in part because such robots are designed to roll on hard flat surfaces. It is difficult to systematically improve such terrestrial robots because we lack understanding of the physics of interaction with complex natural substrates like sand, dirt and tree bark. We are thus limited in our ability to computationally explore designs for potential all-terrain vehicles; in contrast, many of the recent developments in aerial and aquatic vehicles have been enabled by sophisticated computational-dynamics tools that allow such systems to be designed {\em in silico}~\cite{costello2008}.

Compared with human-made devices, organisms such as snakes, lizards, and insects move effectively in nearly all natural environments.  In recent years, scientists and engineers have sought to systematically discover biological principles of movement and implement these in robots \cite{clark2001biomimetic}. This ``bioinspired robotics'' approach~\cite{bhushan2009biomimetics} has proved fruitful to design laboratory robots with new capabilities (new gaits, morphologies, control schemes) including rapid running \cite{clark2001biomimetic,plaAbue06}, slithering \cite{tesch2009parameterized}, flying \cite{ma2013controlled}, and swimming in sand \cite{maladen2009undulatory}. Fewer studies have transferred biological principles into robust field-ready devices~\cite{holAful06,plaAbue06} capable of operating in, and interacting with, natural terrain.

Limbless locomotors like snakes are excellent systems to study to advance real-world all-terrain mobility. Snakes are masters of most terrains: they can move rapidly on land~\cite{jayne1986kinematics,hu09} and through water~\cite{jayne1985swimming}, burrow and swim through sand and soil~\cite{norAkav66}, slither through tiny spaces~\cite{Marvi07112012}, climb complex surfaces~\cite{Jayne2010climbingboa}, and even glide through the air~\cite{socha2002kinematics}. Relative to legged locomotion, limbless locomotion is less studied, and thus broad principles which govern multi-environment movement are lacking. Recently developed limbless robotic platforms~\cite{tesch2009parameterized}, based generally on the snake body plan, are appealing for multi-functional robotics study because they are also capable of a variety of modes of locomotion. These robots can traverse confined spaces, climb trees and pipes, and potentially dive through loose material. However, the gaits that carry these robots across firm ground can be stymied by loose granular materials, collections of particles that display solid-like features but flow beyond critical (yield) stresses. 

\begin{figure}[H]
 \begin{center}
\includegraphics[width = 4.4in,keepaspectratio=true]
 {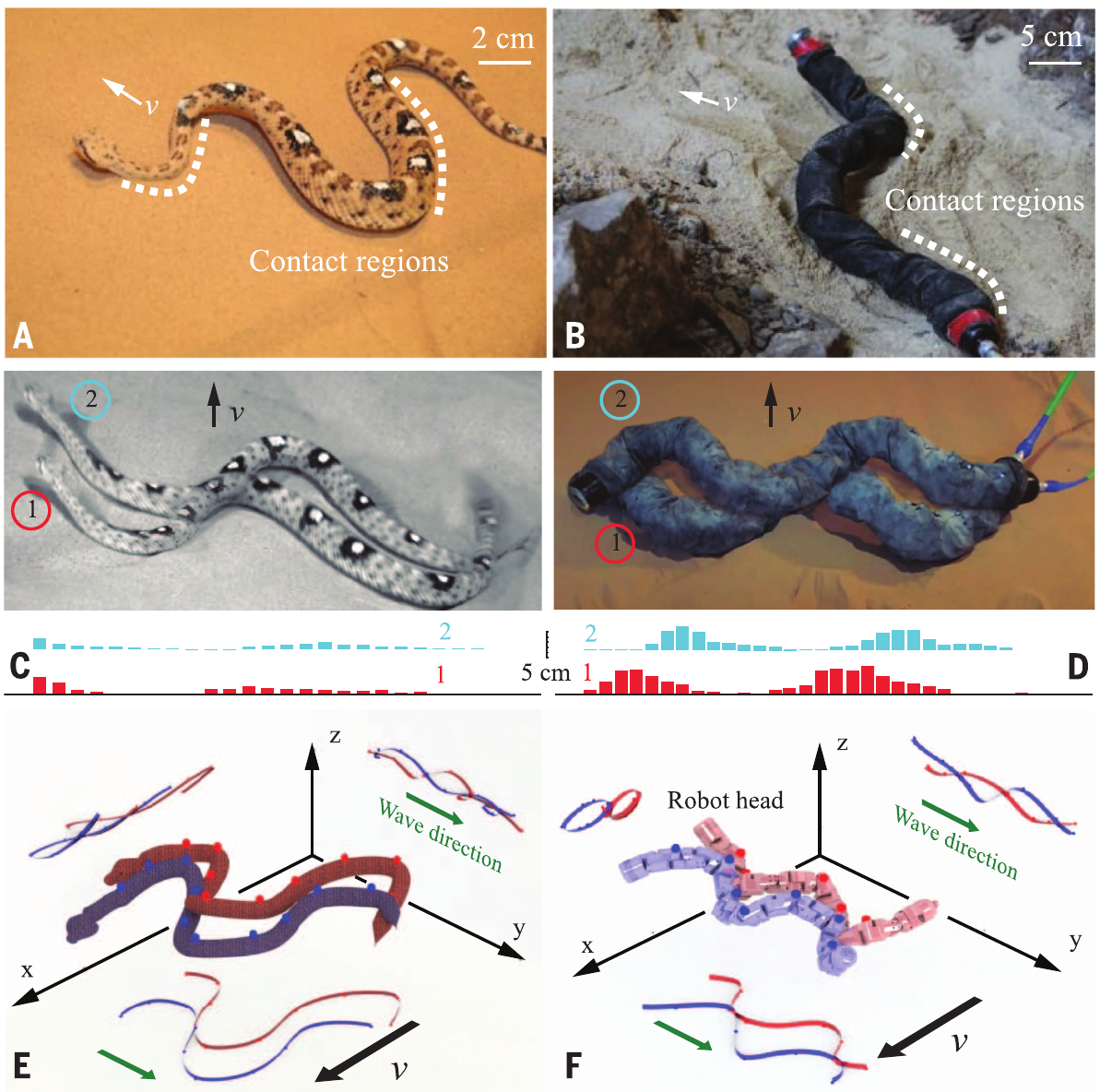}
 \end{center}
 \caption{ Sidewinding locomotion on granular media. (A) A Sidewinder rattlesnake ({\em C. cerastes}) locomoting on a granular substrate of 20$^\circ$ inclination. (B) The CMU modular snake robot traversing granular media at $\approx 20^\circ$ inclination. The arrow indicates the direction of motion. (C and D) Two superimposed snapshots of (C) a sidewinder and (D) the robot moving on level sand. The initial position is denoted by ``1," and the successive position is denoted by ``2." The time between snapshots in (C) and (D) is 0.2 and 3.1 s, respectively. The vertical bars beneath the images indicate vertical displacement along the body at the two times. (E) Two rendered snapshots of a snake sidewinding on level sand, with red indicating the initial time and blue a successive time; time between images is 0.2 s. The 3D configurations were reconstructed according to shape data collected from a sidewinder rattlesnake moving on level sand. Centerline projections of the 3D snake data are shown on three orthogonal planes. The z coordinates of the snake are scaled up four times to increase visibility in the $yz$ and $xz$ projections. The horizontal ($xy$) and vertical ($yz$) waves travel in the posterior direction of the snake with respect to a coordinate system rigidly attached to the animal (as shown by the green arrows). (F) Snapshots of the CMU snake robot executing a sidewinding gait on level sand and projections of the resultant body shapes onto three orthogonal planes. Red and blue indicate the initial and successive images, respectively. Time between images is 6.3 s. The horizontal and vertical waves travel in the posterior direction of the robot with respect to a body-fixed coordinate system (as shown by the green arrows). }
 \label{Fig1}
 \end{figure}

A peculiar gait called sidewinding is observed in a variety of phylogenetically diverse species of snakes  \cite{mosauer1930note, jayne1986kinematics}. Animals with excellent sidewinding ability, like the North American desert-dwelling sidewinder rattlesnake ({\em Crotalus cerastes}) in \fig{Fig1}a,c are often found in environments that are either composed of granular materials (like sand dunes) or are smooth like hardpan desert.  Sidewinding is translation of a limbless system through lifting of body segments while others remain in static (in the world frame, locally rolling/peeling in the frame of the body) contact with the ground. Such interactions minimize shear forces at contact, and thus sidewinding is thought to be an adaptation to yielding or slippery surfaces~\cite{mosauer1930note}. Biological measurements \cite{secor1992locomotor,Marvi_JRSI_Rectilinear} indicate that relative to  lateral undulation, rectilinear locomotion and concertina motion, sidewinding confers energetic advantages on hard surfaces, mainly through lack of slip at the points of contact.  But while sidewinding motion of biological snakes has been extensively studied on hard ground,  \cite{mosauer1930note,gray1946mechanism,jayne1988muscular,secor1992locomotor}, only one study has reported kinematics on granular media \cite{jayne1986kinematics}. 
%; no studies have investigated the mechanics of sidewinding on sandy slopes. Almost nothing is known about the efficacy of this gait on granular media relative to other modes of limbless locomotion.

An initial robotic version of the sidewinding gait has proved useful to enable snake robots to maneuver over flat and bumpy terrain \cite{burdick1994sidewinding}. In a robotic sidewinding mode the device maintains two to three static (locally rolling/peeling) contacts with the substrate at any moment \cite{hatton2010sidewinding}. During this motion, individual segments of the robot are progressively laid into ground contact, peeled up into arch segments, and then transferred forward to become new ground contact segments~\cite{mosauer1930note,hatton2010sidewinding}. However when in field tests a limbless robot (which we will refer to as the Carnegie Mellon University, or CMU robot, see \fig{Fig1}b,d) was confronted with even modest granular inclines ($\approx 10^\circ$) far from the maximum angle of stability ($\approx 30^\circ$), it failed to climb (either slipping or rolling downhill). %Since there had been no detailed study of the mechanics or effectiveness of biological sidewinding on granular substrates (slopes or flat), principles that could improve performance were unknown. 

%%%RESULTS AND DISCUSSION

We posit that the study of success and failure modes in biological snakes can improve the mechanics and control of limbless robot performance, while study of success and failure modes of robots can give insight into important mechanisms which enable locomotion on loose material and perhaps explain why sidewinding has evolved in such organisms. This idea builds on recent work using biologically inspired robots as ``physical models'' of the organisms; such models have revealed principles which govern movement in biological systems, as well as new insights into low-dimensional dynamical systems (see for example ~\cite{holAful06} and references therein). %These insights add a scientific component to the process of bio-inspired engineering.

To discover principles of sidewinding in loose terrain, we challenged {\em Crotalus cerastes} ($N=6$, mass= $98 \pm 18$ g, body-length, tip-to-tail, $L = 48 \pm 6$ cm)~\cite{Mat_supp} to climb loose granular inclines (Table S2 \cite{li2013terradynamics}) of varied incline angle, in a custom trackway housed at Zoo Atlanta (see \figS{FigS2}a). The data comprised 54 trials: 6 snakes, 3 inclinations, 3 trials each. Prior to each experiment, air flow through a porous rigid floor fluidized the granular media and left the surface of the sand in a loosely packed state with a smooth surface; details of this technique are described in~\cite{li2009sensitive}. Three high speed cameras tracked 3D kinematics of 8-10 marked points on the body.  For systematic testing, we chose  three incline angles $\theta  = 0, 10, 20$; the maximum angle of stability $\theta_0$ of the loosely packed material, where local perturbations resulted in system-wide surface flows, was $\theta_0=27^\circ$.

\begin{figure}[H]
 \begin{center}
            \includegraphics[width = 4.5in,keepaspectratio=true]
         {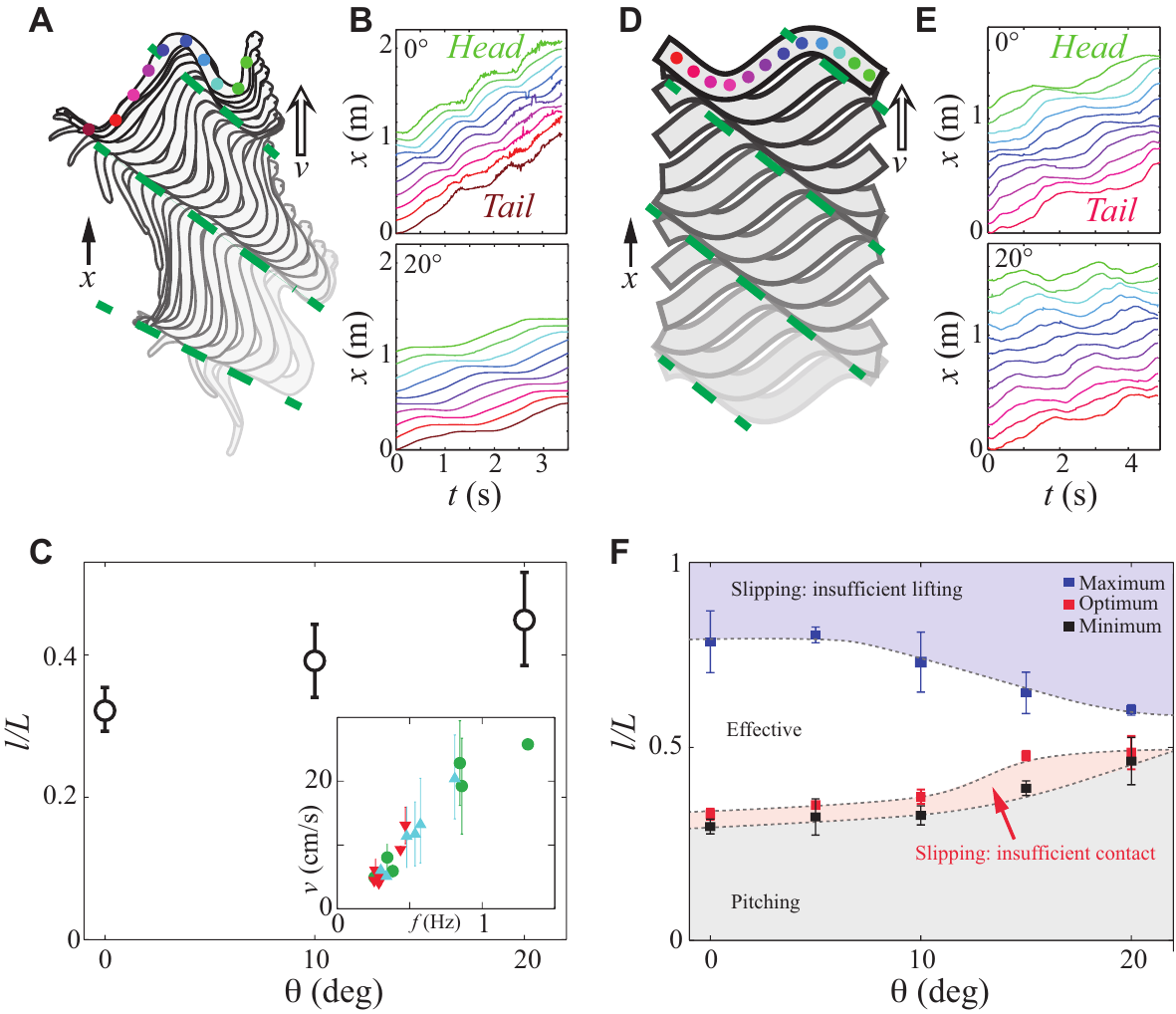}
    \end{center}
 \caption{ Sidewinding motion at different granular incline angles $\theta$. (A) Traces of the sidewinder rattlesnake ({\em C. cerastes}) sidewinding on level sand; green dashed lines indicate tracks that an animal leaves on sand. (B) Displacement versus time of the snake at $\theta=0^\circ$ (top) and $20^\circ$ (bottom); colored lines correspond to colored dots on the snake in (A). Regions of each curve that have zero or near zero slope correspond to static contact with the granular medium. (C) Contact length l normalized by the body length L as a function of $\theta$ for sidewinder rattlesnakes. The data shown encompass six sidewinder rattlesnakes, with three trials per animal at each condition. (Inset) Snake forward speed $v$ versus wave frequency $f$. Green circles, light blue triangles, and red upside down triangles correspond to $\theta=0^\circ$, 10$^\circ$, and 20$^\circ$, respectively. Data denote mean $\pm$ SD. (D) Illustration of the CMU snake robot sidewinding on level sand; green dashed lines indicate the tracks the robot leaves on the sand. (E) Displacement versus time of the robot at inclinations of 0$^\circ$ (top) and $20^\circ$ (bottom); colored lines correspond to the colored dots on the robot in (D). The robot wave frequency in both of these plots was $f = 0.31$ Hz, and its normalized contact length was 0.53 T 0.03 and 0.45 T 0.05 for $\theta=0^\circ$ and 20$^\circ$, respectively. (F) Minimum (black), optimum (red), and maximum (blue) normalized contact length $l/L$ for successful robot climbs as a function of $\theta$. The gray region below the minimum $l/L$ corresponds to pitching failure. The pink region between the minimum and optimum $l/L$ indicates slipping (but still maintains forward progress) due to insufficient contact, and the blue region above the maximum $l/L$ indicates slipping failure due to insufficient lifting. Dashed lines are estimated boundaries of regions of different performance.The robot wave frequency for this plot was $f = 0.08$ Hz. $l/L$ at several other wave frequencies are plotted in Fig. S6B. Data denote mean $\pm$ SD.}
     \label{Fig2}
 \end{figure}

As shown in \fig{Fig1}a,c and \figS{FigS1}a and as previously noted~\cite{mosauer1930note, jayne1986kinematics}, on level granular media the snake generated two contacts (as highlighted by dashed lines in \fig{Fig1}a) with the substrate at each instant, similar to the movement on hard ground \cite{secor1992locomotor,mosauer1930note}. As shown in \fig{Fig1}e, the sidewinding pattern could be approximated by posteriorly propagating waves in both horizontal ($xy$) and vertical ($yz$) planes with a phase offset of $\Delta \phi=1.51 \pm 0.17$ rad.  The position of contact points moved from head to tail, leaving pairs of straight parallel lines on the substrate. The angle between these lines and the direction of motion (\figS{FigS4}d) was $\alpha=33 \pm 8^\circ$ on granular media. The length of each track line was the same as snake body length; the spacing of tracks was thus determined by the spacing of contact points on the snake's body \cite{mosauer1930note} (\figS{FigS4}c). As previously observed on hard ground \cite{secor1992locomotor}, speed increased with frequency (defined here as the number of cycles per second), see inset of \fig{Fig2}c. The snake body intruded into sand by $1.4 \pm 1.3$ cm and created a hill of material at each contact point; for details on penetration depth measurement see supplementary materials and methods \cite{Mat_supp}.

We next examined the behavior of the animals as $\theta$ increased. \fig{FigS1}a and Movie S1 show a sidewinder climbing an inclination of $\theta=20^\circ$. Unlike the robot, the snakes were able to ascend the granular inclines without any axial or lateral slip (Movie S1), indicated by the horizontal regions in \fig{Fig2}b.  In addition, increasing $\theta$ did not significantly change the penetration depth (ANOVA, $F_{2, 51}=0.33, p=0.72$), the angle between direction of motion and track lines (ANOVA, $F_{2, 51}=0.06, p=0.81$), or the number of contacts. Instead, we observed that as $\theta$ increased, the length of the contact regions relative to the body length increased (ANOVA, $F_{2, 51}=48.99, p<0.0001$); we refer to this quantity as $l/L$. To determine this quantity, we analyzed the 3D kinematics data to find the length of snake body in static (peeling) contact with sand in each cycle; see supplementary materials and methods for details on contact length measurement \cite{Mat_supp}. As shown in \fig{Fig2}c, $l/L$ increased by 41\% as $\theta$ increased from $0$ to $20^\circ$. At all $\theta$ portions that were not in contact were lifted clear of the substrate. Wave frequency and climbing speed decreased with increasing $\theta$ (\figS{FigS4}a,b) with effective step length $S_t$ (defined as the distance the head moves forward in each period, see \figS{FigS4}c) remaining constant over all frequencies and $\theta$. For further kinematic details of the movement, see supplementary online text. 

\begin{figure}[H]
    \begin{center}
            \includegraphics[width = 4.5in,keepaspectratio=true]
         {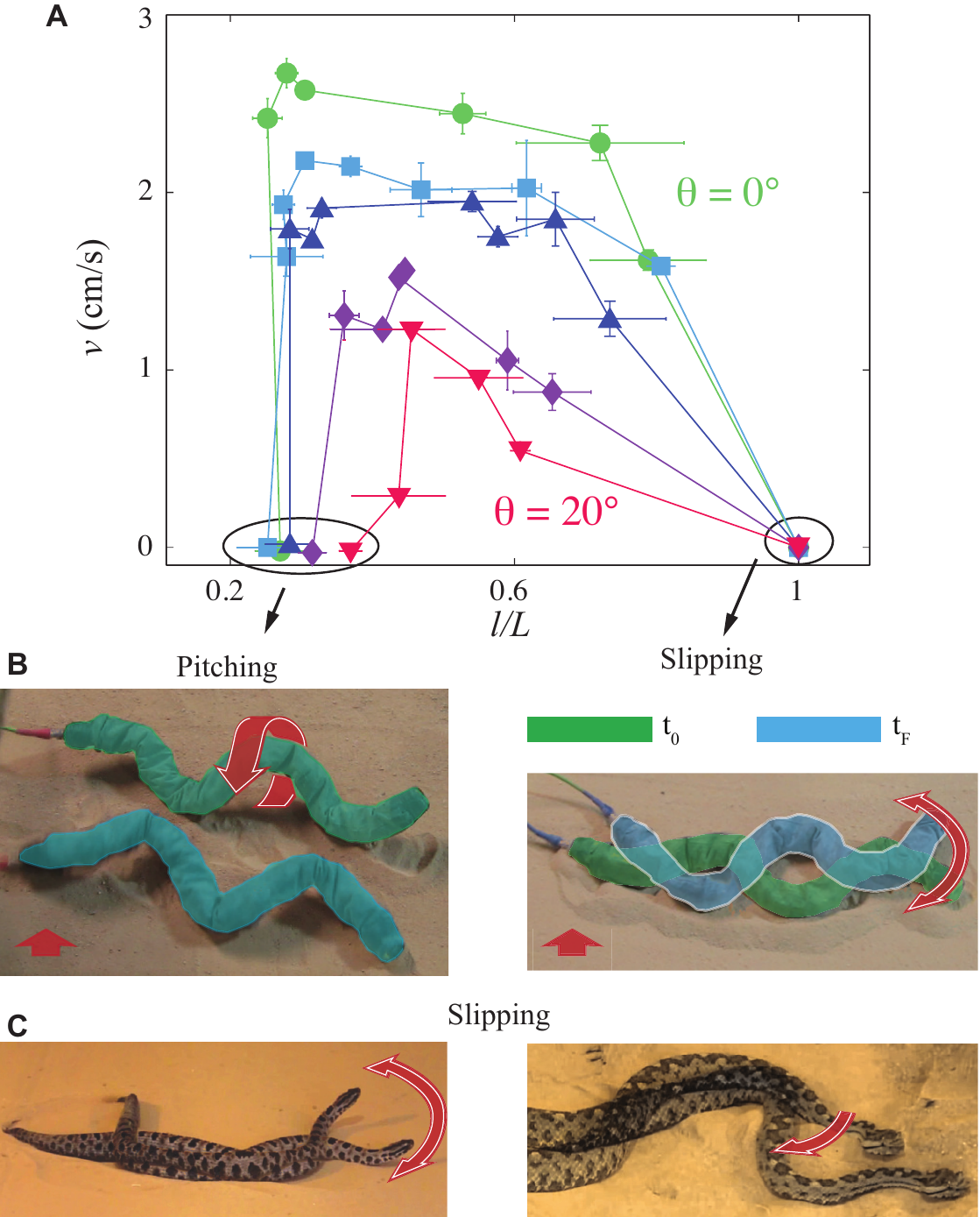}
    \end{center}
 \caption{ Sensitivity of locomotor performance on granular media. (A) CMU snake robot speed versus l/L for inclination angles $\theta=0^\circ$ (green circles), 5$^\circ$ (light blue rectangles), 10$^\circ$ (dark blue triangles), 15$^\circ$ (purple diamonds), and 20$^\circ$ (red upside down triangles). Failure regimes due to pitching and slipping are circled in black. Three trials were performed at each condition. Data indicate mean $\pm$ SD. (B) Super- imposed frames showing pitching and slipping failure modes in the robot ascending $\theta=20^\circ$ and 10$^\circ$ inclines, respectively. Uphill direction is vertically aligned with the page. $t_0$ and $t_F$ represent the time at which each body configuration is captured. The time between two images in the pitching and slipping failure modes is 1.6 and 6.3 s, respectively. (C) Slipping failure of nonsidewinding snakes. Superimposed images show failed lateral undulation on level sand by \textit{S. miliarius} (left, length 47 cm) and failed concertina locomotion on level sand by \textit{M. melanurus} (right, length 47 cm). The time between two images of \textit{S. miliarius} and \textit{M. melanurus} is 0.5 and 9 s, respectively.}
     \label{Fig3}
 \end{figure}

To determine if the ability to climb effectively on sandy inclines is common in closely related snakes, we examined the locomotor behavior of thirteen species of pit vipers (subfamily Crotalinae, Table S1) in the collection of Zoo Atlanta on granular media.  None of these species performed sidewinding in any conditions, instead using either lateral undulation or concertina, sometimes combined with rectilinear, gaits (Table S1).  On level sand, many species failed to achieve forward movement, and on sand slopes at $\theta=10^\circ$, all but a single species failed to move uphill (\fig{Fig3}c and Supplementary Movie S3).  %On level surfaces, both lateral undulation and slide-push locomotion resulted in both failures and successes, with different species achieving different outcomes from the same locomotor mode (Table S1).  
Yielding of the sand was accentuated by the failure of these species to lift portions of their body above the substrate (Supplementary Movie S3). 

%%%%%%% SNAKE ROBOT SECTION %%%%

We hypothesize that the sidewinder rattlesnake's ability to move effectively on loose substrates (compared to the robot and the other closely related snakes) is made possible by neuromechanical control that generates appropriate contact length as sandy inclines become more susceptible to flow. To test this hypothesis, we next used the CMU robot as a physical model to study locomotor performance as a function of $l/L$ for different $\theta$~\cite{Mat_supp}. In particular, the CMU robot's ability to deform in arbitrary modes (using 17 modules, see Section 1.3 in ~\cite{Mat_supp}), including a combination of traveling waves which generate sidewinding, makes it an attractive model on which to test the generality and efficacy of the contact length mechanism. And despite the differences in weight and size of the snake robot relative to the organism, as demonstrated in previous studies of fish, turtles, and cockroaches~\cite{sefati2013mutually,mazouchova2013flipper,holAful06} (and shown below), principles of small organism locomotor bio and neuromechanics can be deduced through systematic variation of parameters in larger-scale physical models.

To generate sidewinding in the robot, we controlled its modules (and joints) to generate two posteriorly-directed traveling waves in the horizontal and vertical planes with a 1/4 period phase offset, $\Delta \phi=1.57$ rad (comparable to that of the animal).  These waves were generated such that dorsal and ventral surfaces maintained their respective orientations throughout the motion. This produced a pattern of undulation and lifting similar to that observed in the biological snakes (\fig{Fig1}e,f). When the traveling waves are approximated using sinusoids, the resultant shape of the robot resembles an elliptical helix whose minor axis is perpendicular to the ground and its major axis is aligned with the direction of motion. Similar to the biological snake, during a gait cycle each module (segment) traced an approximately elliptical trajectory (\fig{Fig1}f). All experiments were performed at constant wavelength ($\lambda=0.5L$) and at frequencies sufficiently low as to avoid motor angular position and speed error.  On level hard ground, a minor to major axis aspect ratio of $0.9$ produced steady motion at a speed of $0.03$ m/s (at frequency $f=0.08$ Hz). On hard surfaces at $\theta=20^\circ$, decreasing the aspect ratio to $0.7$ allowed the robot to climb without pitching or rolling down the hill, and at comparable speeds as on level hard ground~\cite{hatton2010sidewinding}. 

The waves that generated effective robot sidewinding on solid inclines were ineffective on granular media; see Movie S2.  To improve the performance we decreased the aspect ratio of the elliptical core, which increased the contact length in a manner analogous to the behavior observed in the biological snakes (see \fig{FigS6}a). The minimum $l/L$ needed for successful climbing (defined as ascending at positive forward speed without pitching, see example traces in \fig{Fig2}d and tracked markers in \fig{Fig2}e) increased by approximately 70\% as $\theta$ increased from $0$ to $20^\circ$. Within the successful regions, we were further able to optimize the contact lengths to minimize slip and slightly improve the speed as compared to moving with the lower-bound of $l/L$.  For details on robot contact length measurements refer to \cite{Mat_supp}. There was no dependence on frequency, indicating that body inertial forces were minimal and that granular frictional forces determined the resistive forces~\cite{albert1999slow}. Thus despite the many differences between the biological and robot sidewinding, comparable performance was achieved through use of a similar strategy of increasing contact length with increases in $\theta$. We note that the elliptical helical gait is only one way to generate the traveling wave that propels the sidewinding in both hard ground and sand. Experimenting with helices of other cross-sections (such as oval shapes) led to similar results.  We also note that the maximum penetration depth $d$ at the minimum $l/L$ for successful climbing (at $f=0.08$ Hz) decreased by  $\approx$ 30\% as $\theta$ increased from $0$ to $20^\circ$ (ANOVA, $F_{4, 10}=9.03, p=0.007$).

We next systematically studied how robot performance changed with $l/L$, see \fig{Fig3}a. As shown in \fig{Fig2}f and \fig{Fig3}a, the robot was able to ascend effectively for a given $\theta$ only within a range of $l/L$. The width of the range narrowed as $\theta$ increased, indicating that for shallow slope angles, sidewinding performance was robust to variations in contact length, while for higher slope angles, the control effort must increase to target a narrower range of contact length. On inclinations of less than $15^\circ$, the robot exhibited high performance (speed) in a wide range of $l/L$ until too large $l/L$ produced slipping that decreased forward speed; this was related to an inability of the robot to lift itself sufficiently to avoid unnecessary ground contact (similar to the slipping failures observed in other snake species). At the other extreme, small $l/L$ resulted in insufficient supporting region, causing the robot to pitch down the slope. In the pink region between minimum and optimum $l/L$ in Fig. 2F, robot slipped due to insufficient contact but still made forward progress.  For large $\theta$, even within the range of effective ascent, performance was degraded due to downhill slip, which decreased the effective step length (see \figS{FigS6}d). 

%%%%%%% GRANULAR DRAG SECTION %%%%

The success of the robot and its similar performance to the biological sidewinder suggest the following picture of sidewinding on sandy slopes. First, the animal targets a neuromechanical control scheme consisting of two independently controlled and appropriately phased orthogonal waves (Fig. 1E,F). We note that such a scheme satisfies the definition of a ``template'' -- that is a behavior ``contains the smallest number of variables and parameters that exhibits a behavior of interest''~\cite{holAful06}. Discovery of templates is useful as they provide an organizing principle for the enormous number of degrees of freedom inherent in all organisms. Second, the robot experiments suggest that the two wave template dynamics can be simply modified on sandy slopes such that the animal targets a pattern of $l/L$ and body-segment lifting which minimizes slip and pitching.

Understanding the ground reaction forces responsible for the relevant interactions (e.g. optimal $l/L$ without slip) requires a model of the interaction of objects with granular media. However, there is no fundamental theory for bio- and robotically-relevant interactions in granular media and a wider class of materials (mud, rubble, leaf litter). In particular, despite recent discoveries in terradynamics \cite{li2013terradynamics}, drag \cite{albert1999slow}, and impact \cite{katsuragi2007unified} on level dry granular media (and theoretical approaches to characterize granular rheology \cite{henann2013predictive}), none of these studies have investigated drag forces on granular inclines.

We therefore studied the physics of transient granular yield forces as a function of $\theta$. For context, we note that studies of steady state drag on level granular surfaces reveal that force is proportional to plate width, $w$, and depth squared, $d^2$ \cite{costantino2011low}. To make the first measurements of drag force $F$ dependence on $\theta, d$ and $w$, we performed drag force measurements on granular inclines \cite{Mat_supp}. We inserted flat plates of different widths ($w=1.5-6$ cm) to different depths ($d = 1-2$ cm) into the granular medium perpendicular to the slope plane, and displaced the plate downhill $15$ cm. Depths were chosen to bound the range of penetration depths observed in the animal experiments. Because we were interested in the yield behavior of the granular slope in the quasi-static limit (we assume inertial effects are small at speeds at which the snakes generally operate) we performed drag at slow speeds and studied the force increases under small displacements $\delta$ from rest.

\begin{figure}[H]
    \begin{center}
            \includegraphics[width = 4.5in,keepaspectratio=true]
         {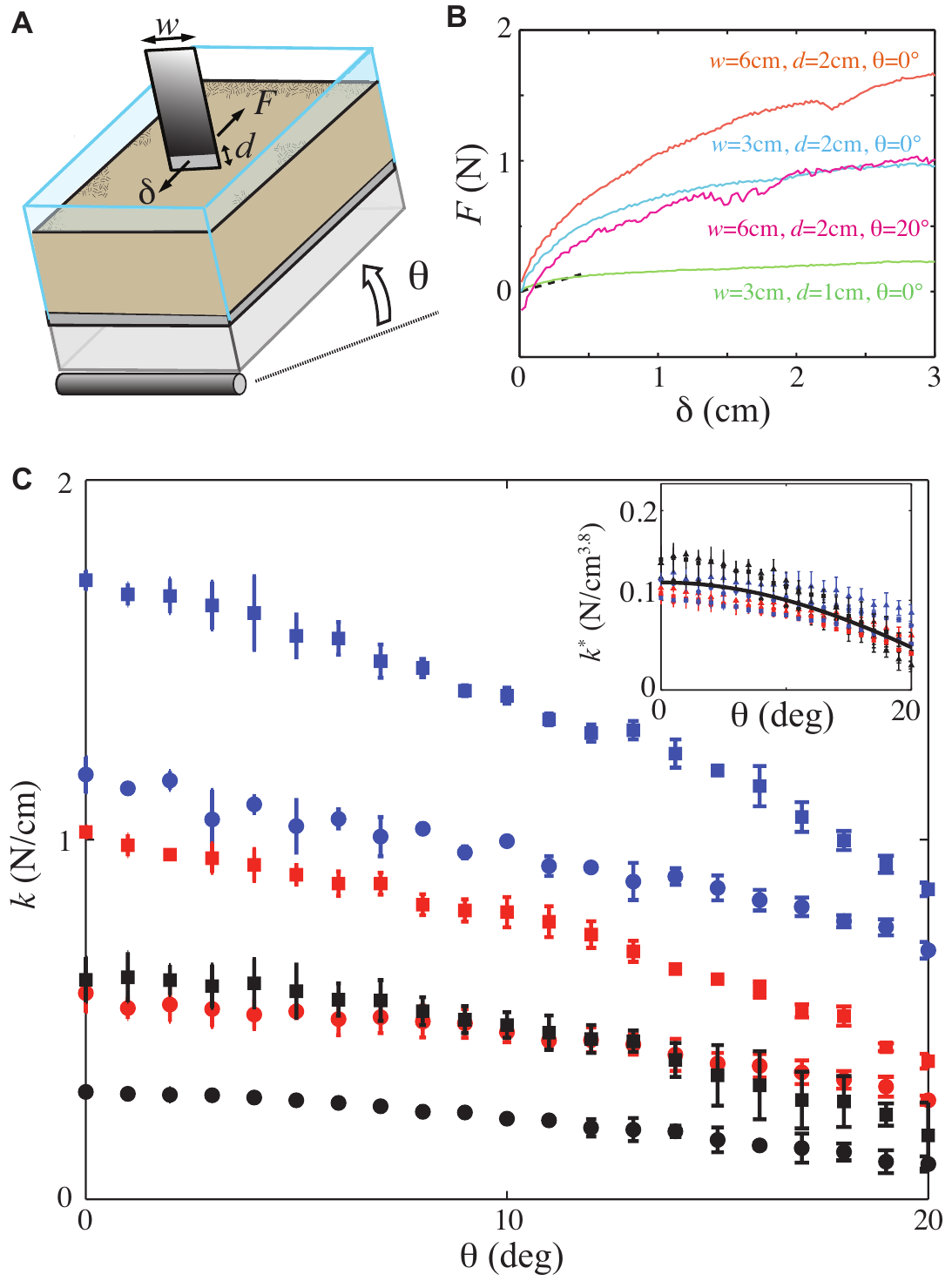}
    \end{center}
 \caption{Granular incline drag force experiments. (A) Schematic of drag force apparatus. (B) Drag force, $F$, versus horizontal displacement, $\delta$, for different plate widths, $w$, and penetration depths, $d$. The dashed line indicates the region used to calculate sand stiffness ($k$). (C) The fitted $k$ versus inclination angle, $\theta$, for different $w$ and $d$. The inset shows the sand stiffness data collapse when normalized by $w^{0.8}d^{2}$; the thick black line is the fit function $0.12 \cos (\frac{\pi}{2\theta_0} \theta)$ with $\theta_0=0.47$ rad ($\theta_0=27^{\circ}$). In the main panel and the inset triangles, circles, and squares represent $w=1.5, 3, 6$ cm, respectively. Black, red, and blue colors illustrate $d=1, 1.5, d=2$ cm, respectively. Data denote mean $\pm$ SD. }
     \label{Fig4}
 \end{figure}

As shown in \fig{Fig4}b, force increased substantially for small displacements until it reached a saturation regime, after which there was a slower increase in force associated with the buildup of a granular pile in front of the plate. We refer to the regime of rapid rise as the ``stiffness'' of the sand (before significant yielding) and denote it $k$ ($F/\delta$ near $\delta=0$). We estimated $k$ by fitting a line to the first 17\% of data (corresponding to a displacement  $0.5$ cm, before material yielded significantly) presented in \fig{Fig4}b, and plot this in \fig{Fig4}c. For fixed $d$, $k$ decreased by $\approx 50\%$ as $\theta$ increased from $0$ to $20^{\circ}$ (\fig{Fig4}c). For a fixed $\theta$, $k$ increased quadratically with $d$ and nearly linearly with $w$. This scaling indicates that in the limit of shallow drag and small displacements $\delta$, $F \propto k \delta$ with $k \propto w^{0.8} d^{2}\cos (\frac{\pi}{2\theta_0} \theta)$ where the phenomenological $\cos$ term reflects the rapid decrease in force as $\theta$ approaches $\theta_0$ (see the supplementary materials and methods for details on curve fitting \cite{Mat_supp}).  We note that for $\theta=0$ and fixed $\delta$, the dependence on $w$ is similar to that previously observed \cite{costantino2011low} even though our depths are shallower.

The drag force measurements support our hypothesis of contact length control: as $\theta$ increases, effective sidewinding can be maintained by increasing the contact length to offset the decrease in yield forces. This allows the locomotor to maintain stresses below the yield stress, thereby minimizing slip. Insights from the drag measurements also indicate why the appropriate amount of body lift and contact to be within the range of $l/L$ is important. If lift is too small, other segments of the locomotor encounter drag forces which must then be offset by either increases in $l/L$, or potentially increasing intrusion depth. However, increases in $l/L$ would decrease lift, resulting in greater drag. We note that increasing contact length relative to increasing intrusion depth has obvious benefits because intrusion into granular media requires yielding material and the force to do so increases with depth (see the challenges associated with penetration in~\cite{Winter2012}). The energetic cost to vary $l/L$ is small in comparison. 

The contact length modulation strategy has benefits in terms of locomotor control: once the animal or robot is moving using a sidewinding template (again, two independently controlled waves with phase difference of approximately $\pi/2$) slip and pitch mitigation in flowable substrates can be effected with relatively simple modulations of the basic template wave pattern. We note that this control has features in common with our previous biological~\cite{mazouchova2010utilization} and robotic \cite{mazouchova2013flipper,li2009sensitive} studies of relatively slow legged movement on the surface of granular media, which also demonstrated that use of the solid features of granular media had benefits in certain locomotor regimes. We predict that other locomotors that move on granular surfaces could target movement patterns whose modulation can be used to achieve minimal-slip.

As expected, there were differences in the details between the sidewinding in the two systems. We view these differences, necessarily present due to the relative simplicity of the robot compared to the organism, as consequences of the different mechanisms which are used (biological complexity vs relative robot simplicity) as ``anchors'' of the templates, in the terminology discussed in ~\cite{holAful06}. For example, the CMU robot has no compliance in its joints, many fewer degrees of freedom, and larger range of rotation (-90 to 90 degrees) compared to biological snakes. Another interesting difference between a snake and the robot is that although both of them slowed as $\theta$ increased (see Fig. S6c and Fig. S4a), the snake's decrease in speed was correlated with a decrease in frequency (indicating possibly either muscle limitations or active control to decrease inertial forces), whereas the robot speed decrease was largely determined by a decrease in spacing between tracks as shown in Fig. S6d. This decrease in effective step length was related to slipping of the robot at the highest $\theta$. Comparative study of the anchoring mechanics is useful to learn about what lower-level mechanisms in the control hierarchy are critical, both to generate template dynamics as well as to understand neuromechanical control targets for the anchors.

%We close by proposing that further investigations of the type we illustrate here can give insight more broadly into how fundamental locomotor principles emerge~\cite{Anderson1972} in complex biological (and increasingly robotic) systems and requires the integration of many disciplines (e.g. physics, biology, mathematics, engineering). Such integration revives the original spirit of cybernetics promoted by Wiener~\cite{wienerbook} over 60 years ago, in that it emphasizes and adds locomotion to the important aspects of ``control and communication in the organism and machine'' (see also~\cite{cowan2014feedback,roth2014comparative} for recent discussion of similar issues). Robotic locomotors which can coexist alongside animal counterparts can fulfill the dream of one of the early cyberneticians, Louis Couffignal, who defined the field as  ``the art of ensuring the efficacy of action''~\cite{coufarticle}. What makes this new is that as studies of aerial and aquatic locomotion have benefited from fundamental studies of hydrodynamics, future work on terrestrial locomotion will likely benefit from increased understanding of the physics of soft materials. 

\bibliography{snakebib}

\bibliographystyle{Science}

%\begin{scilastnote}

%\item 
\noindent {\bf Acknowledgments:}
We thank Matt Tesch, Ellen Cappo, Justine Rembisz, and Lu Li from Robotics Institute at Carnegie Mellon University for helping with the robot experiments; Jason Brock, David Brothers, Robert  Hill, Luke Wyrwich, and Brad Lock from Zoo Atlanta for helping with snake experiments; A. Young and K. Young for assistance with collecting snakes; Dante Dimenichi, Robert Chrystal, and Jason Shieh from Georgia Tech for helping with construction and video analysis; Tim Nowak and Candler Hobbs for photography; Vadim Linevich for CAD drawings; Paul Umbanhowar and Andrew Zangwill for helpful discussion; the National Science Foundation (CMMI-1000389, PHY-0848894, PHY-1205878, PHY-1150760), Army Research Office under grants W911NF-11-1-0514 and W911NF1310092, the Army Research Lab MAST CTA under grant number W911NF-08-2-0004, the Elizabeth Smithgall Watts endowment, for financial support. D. I. G., H.C., and D.H. also acknowledge ARO and NSF PoLS for supporting the Locomotion Systems Science Workshop in Arlington, VA, May 2012.\\

%\end{scilastnote}

\noindent {\bf Supplementary Materials} \\
www.sciencemag.org\\
Materials and Methods\\
Figures S1, S2, S3, S4, S5, S6\\
Tables S1, S2\\
References (34-38) \\ 
Movies S1, S2, S3, S4, S5

%%%%%%%%%%%%%%%%%%%%%%%%%%%%%%%%%%%%%%%%%%%%%%%%%%%%%%%%%%%

%\clearpage
% http://commons.wikimedia.org/wiki/File:Dunes_at_Death_Valley.jpg

\newpage
 
 \begin{center}
{\bf {\large Supplementary Materials for}} \\
Sidewinding with minimal slip: Snake and robot ascent of sandy slopes\\
Hamidreza Marvi, Chaohui Gong, Nick Gravish, Henry Astley, \\ \ Matthew Travers, Ross L. Hatton, Joseph R. Mendelson III, Howie Choset, \\
David L. Hu \& Daniel I. Goldman\\
Correspondence to:  daniel.goldman@physics.gatech.edu

\end{center}
%%%%%Methods%%%%%%%%%

{\bf This PDF file includes:}

Materials and Methods

SupplementaryText

Figs. S1 to S6

Tables S1 and S2

Captions for Movies S1 to S5

References 34-38

Other Supplementary Materials for this manuscript includes the following: 

Movies S1 to S5

 \newpage

\section*{Materials and Methods}

\subsection*{Snakes, fluidized bed, visualization, and tracking}
Data was collected for six adult sidewinder rattlesnakes, {\em Crotalus cerastes} ($N = 6$, $m = 98 \pm 18 $ g, $L = 48 \pm 6$ cm) and 200 kg of sand from Yuma County, Arizona (\figS{FigS1}a. All of these snakes were housed at Zoo Atlanta where the experiments were conducted. We designed and built an air-fluidized bed (\figS{FigS2}a) of size 2 x 1 m$^2$ to control the sand compactness 
and inclination angle.  In the fluidized bed, above a critical air flow rate (against gravity), the granular material behaves like a fluid. Gradually decreasing the flow to zero results in the media reaching a loosely packed state with a flat surface; previous disturbances to the medium are no longer present.

We recorded kinematic shape change data (snake and robots) using 3 AOS X-PRI high speed cameras at 120 fps. We calibrated the cameras, analyzed all of the videos, and obtained the position of each dot on snake body as a function of time using an automated image processing software developed in MATLAB \cite{hedrick2008software}. We used a calibration object with 49 markers (\figS{FigS2}b) to calibrate our 3 cameras. As detailed in the {\bf Contact length measurement} section, we used the 3D shape data to find the portion of the animal's body in contact with sand. Penetration depth was obtained based on the z-displacement of the dots on snake body; see \S~\ref{Penetration depth measurement} for details. Body speed was measured as distance traveled per unit of time averaged over all points tracked on the snake's body; regularization was performed such that contributing measurements were taken from periods where the average gait frequency varied by less than 10\%. The wave frequency was found by analyzing the x-displacement data using a peak detection algorithm \cite{Marvi_JRSI_Rectilinear}.  To determine the phase offset between $xy$ and $yz$ waves we found the phase of horizontal wave ($xy$) at the static regions of body. We know the phase of vertical wave ($yz$) at the center of these static regions since they correspond to the lowest point of a sine wave. We assumed a sinusoidal curvature and determined $xy$ phase of the static regions based on the body curvature and its rate of change at these points. The mean of these $xy$ phase values was the center of the static region, corresponding to the lowest point in the vertical wave.

\subsubsection*{Contact length measurement}
\label{contact length measurement}

We provide a more detailed description of our methods for measuring contact length of the snake and snake robot. The MATLAB program DLTdv4~ \cite{hedrick2008software} was used to analyze synchronized videos, recorded using 3 AOS high speed cameras, to obtain time resolved $xyz$ positions of tracking dots on the dorsal surface of the snake's body (Fig. 2b,e). We tested the accuracy of this method by putting 5mm spherical markers at known $xyz$ positions on sand and computing their positions in the tracking software. The $xyz$ coordinates of these markers were determined to within 2mm (the penetration depth of snakes were on the order of $1.4 \pm 1.3$ cm). Although the snakes generate continuous traveling waves as they move, parts of the body form regions in static contact with the substrate (unless the material yields). These instantaneously static segments roll/peel away from the substrate as the snake continues to move. This process is is similar to a slip-free wheel, which makes instantaneous static contact with the ground as it rolls. The static contact is well illustrated by observation of snake ventral scale imprints on the sand(see \figS{FigS2}c). 

Thus, to obtain the normalized contact length for the snakes, we need to find the proportion of the snake body that has zero velocity at each instant. Using our digitally tracked 3D points we determined snake-ground contacts by examining the velocity profile along the body of the snake.  We calculated the numerical time derivative of displacement in the direction of travel for each dot on the snake's body. We then found the proportion of time this derivative was below a manually tuned threshold in each period (\fig{Fig2}b,e). We averaged this proportion for all of the dots on each snake in each trial. We report this average as the proportion of time the snake was in contact with the ground in each trial.  According to the following discussions this parameter is equal to the portion of snake length that is in contact with sand, which we call the normalized contact length $l/L$.   

To find the contact segments for the snake robot we stopped the robot at the same phase after each trial and directly measured the robot-sand contact length. This contact length was independent of the phase at which we stopped the robot (all experiments were performed at constant wavelength ($\lambda=0.5L$)). This direct measurement removed the need for 3D data analysis.

In addition, we used video inspection to validate our contact length calculations. We used Open source Tracker, a Video Analysis Tool (\url{www.cabrillo.edu/~dbrown/tracker/}), to manually analyze each video and estimate the portion of snake and robot body that was in contact with the sand in each trial. The manual video inspection proceeded in the following steps: We opened up the video of interest in Open source tracker. We then chose a configuration/phase of the animal that is easy to visualize the segments that make contact with sand; see \figS{FigS1} (we consistently used the same configuration for all of the videos, although contact length remains constant during each period). We then manually measure the length of body in contact relative to the entire length of the animal. Regions of the snake body in which a visual shadow below the body were observed were classified as the lifted segments of the body, the portions of the snake body which were in contact with the sand did not have a shadow and were often signified by a lack body curvature in these regions (dashed red lines in \figS{FigS1}). The equally spaced dots we put on snake body helped us better estimate the normalized contact length. As shown in \figS{FigS1}, the contact regions were clear from the side view and were consistent between different trials (small error bars in \figS{FigS3}a). As shown in \figS{FigS3}a,b the results of the manual video inspection are consistent with our previously reported data for contact length of snake/robot.

We assume wave speed and contact length stay constant in a gait cycle and any point on the substrate is touched only once by the snake body. As the wave travels down the snake body, the amount of time for a particular marker being static, $t_s$ is proportional to the length of contact,
\begin{equation}
\frac{l}{L}=\frac{t_{s}}{T}.
\end{equation}
As discussed above, we find $t_{s}/T$ (the length of horizontal segments of the plots in \fig{Fig2}b divided by the period $T$) and average that over all of the markers to calculate the normalized contact length $l/L$.

\subsubsection*{Penetration depth measurement}
\label{Penetration depth measurement}

We found penetration depth of snakes using the z-displacement of the markers on snake's body using the visual tracking system discussed above. We assumed a constant dorsoventral thickness for the snakes and searched for the maximum $z$-displacement of the markers in each trial and defined this as penetration depth. However, we needed to know where the sand surface was relative to an inertial frame to get accurate $xyz$ coordinates of the markers. Thus, before each trial we fluidized the bed to obtain a uniformly flat granular surface with minimum variation from trial to trial.  Our experimental test bed was constructed specifically for this purpose, with a porous floor along the entire bottom area so uniform air flow through the sand is possible. This resulted in a uniformly reconstructable granular surface being created between each trial. The air fluidization technique has been used in studies of locomotion on granular media to repeatedly reconstruct near identical environmental initial conditions \cite{maladen2009undulatory,li2009sensitive}. 

\subsection*{Granular}

Granular drag experiments were performed in an air fluidized bed of length, $L=43$~cm, and width, $W=28$~cm, filled with the same sand used in the snake experiments. Air flow through a porous floor in the bed fluidized the sand into a loosely packed granular state consistent with that used in locomotion experiments. After fluidization and once the sand settled, we slowly (0.9 degrees/s) tilted the bed to an angle between $0-20$ degrees. A flat plate of width, $w=1.5-6$cm, was inserted to a depth $d = 1-2$~cm and displaced a distance of 6~cm at a speed of 0.65~cm/s along the sand slope. Plate displacement was performed by a Firgelli L16 miniature linear actuator. The flat plate was constructed of 1~mm thick aluminum; four strain gauges (Omega KFG) were attached to the top of the plate. Using the strain measurements, we were able to extract the granular drag force experienced by the flat plate. We used MATLAB curve fitting and optimization tool boxes to find the best fit of the measured stiffness data.  Our choice of fitting function was based on the physical characteristics of the granular media. We first collapsed the stiffness data by dividing it by $d^a w^b$ \cite{costantino2011low}.  At each inclination angle, $theta$, $k$ increased quadratically with $d$ and nearly linearly with $w$ (similar to {\em steady state} drag on level granular surfaces \cite{costantino2011low}). We then fit the function $c \cos (\frac{\pi}{2\theta_0} \theta)$ to the reduced data. The $\cos$ term reflects the rapid decrease in force as $\theta$ approaches $\theta_0$. 

\subsection*{Robot}

The CMU modular snake robot was developed at Choset's lab ($m$ = 3150 g, $L$ = 94 cm) and was used for our robotic experiments (\figS{FigS5}c,d) \cite{wolf2003mobile,wright2007design,hatton2010sidewinding}. This robot is composed of 16 identical modules. Each module is 5 cm in diameter and can output over 3 Nm of torque.  Each of these modules serves as a single-DOF joint and the joints are chained together such that they rotate alternatively in pitch and yaw directions with respect to a body-fixed frame.  This mechanical configuration enables the snake robot to form three-dimensional shapes. The robot's tail module includes a tether connection with integrated slip-ring.

%%%%%%%%%%%%%%%%%%%%%%%%%%%%%%%%%%%%%%%%%%%%%%%%%%%%%%%%%%%
\section*{Supplementary Text}

\figS{FigS1} shows a sidewinder rattlesnake and CMU robot traversing loose sand at inclination of $20$ degrees. The lifting regions for both snake and robot are highlighted by red bars. 
 
As shown in \figS{FigS4}c spacing of tracks $S_t$ did not change significantly with incline angle (ANOVA, $F_{2, 51}=0.73, p=0.49$).  However, wave frequency decreased with increasing $\theta$ (\figS{FigS4}b, ANOVA, $F_{2, 51}=5.93, p=0.005$). The decrease in wave frequency led to a decreased forward speed with increasing incline angle (\figS{FigS4}a, ANOVA, $F_{2, 51}=7.84, p=0.001$). As shown in \figS{FigS4}d, the angle between snake tracks and the direction of motion was also $\alpha=33\pm8$ degrees on granular media and did not significantly change across inclines (ANOVA, $F_{2, 51}=0.09, p=0.77$). Open source Tracker, a Video Analysis Tool (\url{www.cabrillo.edu/~dbrown/tracker/}), was used to manually analyze each video and find $S_t$ and $\alpha$. 
 
Choset's group at CMU has developed a family of limbless robots called modular snake robots \cite{wolf2003mobile,wright2007design,hatton2010sidewinding}. These robots comprise $16$ independent modules, each with a single joint arranged orthogonal to the previous one. A schematic of sidewinding gait, sidewinding backbone, and CMU robot schematic are shown in \figS{FigS5}. 

Normalized contact length of the robot is plotted versus aspect ratio in \figS{FigS6}a. The plot shows only the minimum required contact lengths for successful climbing. As we increase aspect ratio, the vertical wave amplitude increases. As a result of the increased lifting action, the normalized contact length $l/L$ decreases. With increasing $\theta$ we had to decrease the aspect ratio of the elliptical helix for successful climbing (corresponding to an increase in minimum required $l/L$). The minimum required $l/L$ for successful climbing of the CMU robot is plotted versus $\theta$ in \figS{FigS6}b. The wave frequency did not have an impact on $l/L$ within the range of frequencies the robot can handle without significant joint error.  As shown in \figS{FigS6}c, speed of the robot at each frequency decreased with increasing $\theta$. Speed also depended on frequency and the higher the frequency, the higher the speed. The reason speed of the robot was decreased with increasing $\theta$ at each frequency (\figS{FigS6}c) was the decrease in spacing of tracks at higher inclinations (\figS{FigS6}d). As shown in Movie S1, there was more local slipping at higher inclinations (even in the range we call effective locomotion on \fig{Fig2}f). Thus, the spacing of the tracks was reduced by increasing the inclination. 

Table S1 presents locomotor performance of thirteen species of pit vipers on sand at zero and ten degree inclinations. These results show that while several viper (and non-viper) species can sidewind \cite{jayne1986kinematics,gans1972sidewinding}, this ability is neither universal nor even common.  Further examination and broader taxonomic sampling is the subject of future work.

%The change of aspect ratio $p$ results in the change of contact ratio $l/L$ and COM height $h$.  When the aspect ratio varies from 0.0 to 0.4, $l/L$ can be closely approximated by the linear relationship $l/L=1-(7/3) p$ (determined from experiment data). According to the kinematic model, the COM height can be computed as $h=54.53 p=23.37 (1-l/L)$. For the robot to rest statically on a slope of inclination $\theta$, the width of the supporting base should be $w/2 \ge \tan(\theta) h$, where $\theta$ is the slope angle, and $w=0.5 l$ (there are two contacts with the substrate at each instant).
%Thus, $\theta$ could be calculated according to \eq{eq_4}. 
%\begin{equation}
%\theta=\arctan(l/(4 23.38 (1-l/L))). 
%\label{eq_4}
%\end{equation}
%\figS{FigS7} shows the experimental (blue) and theoretical (red) pitching contact length as a function of inclination angle $\theta$. 

 \newpage

    \begin{suppfig}[H]
    \begin{center}
            \includegraphics[width = 6.5in,keepaspectratio=true]
         {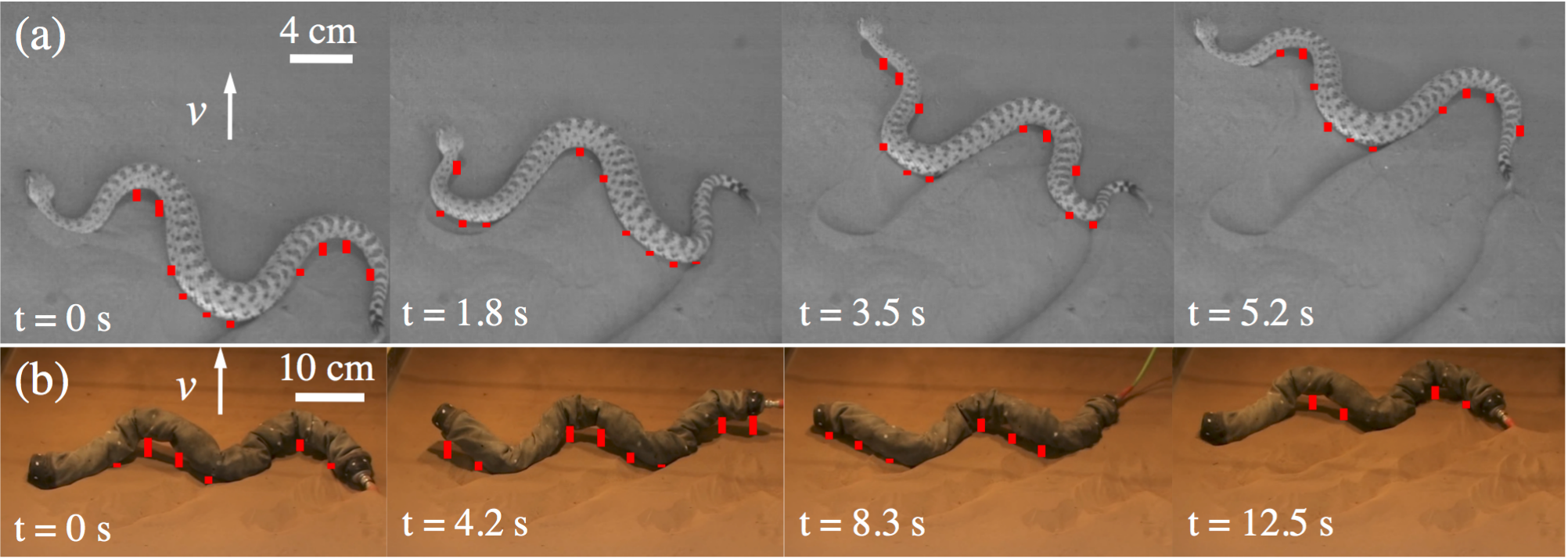}
    \end{center}
 \caption{(a) A sidewinder rattlesnake and (b) CMU snake robot traversing a 20 degree granular slope. The lifting regions are highlighted by red bars. }
     \label{FigS1}
 \end{suppfig}
 
  \newpage
 
    \begin{suppfig}[H]
    \begin{center}
            \includegraphics[width = 6.5in,keepaspectratio=true]
         {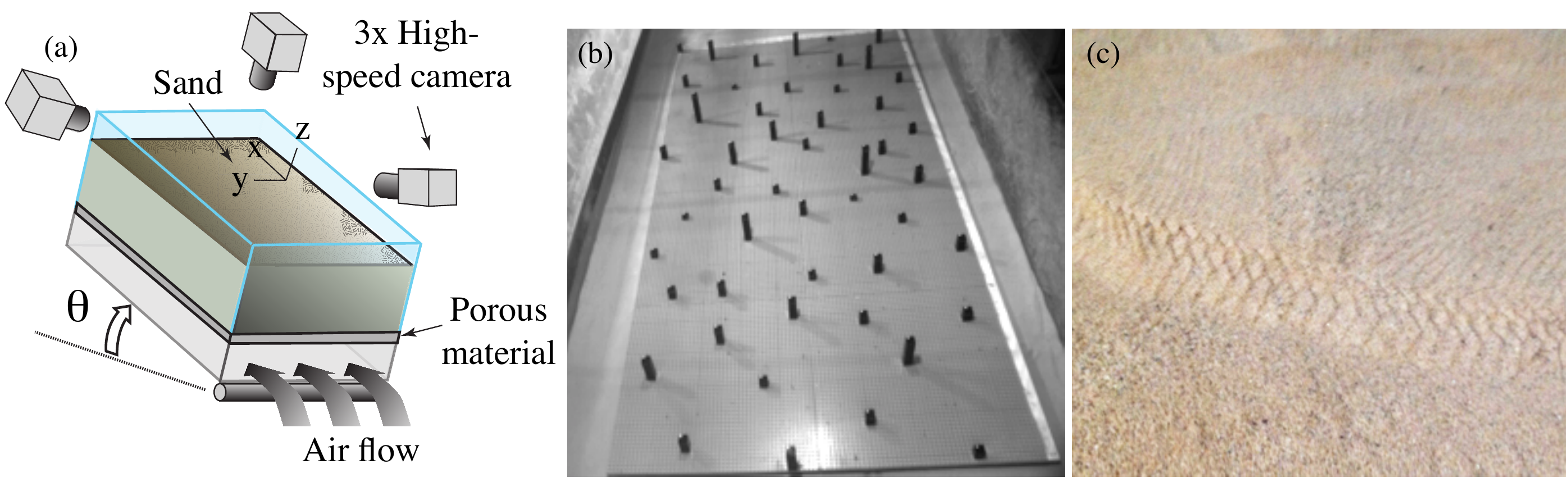}
    \end{center}
 \caption{(a) The experimental apparatus including a diagram of a tiltable fluidized bed trackway and the 3D high-speed imaging system.  (b) The calibration object covering the entire bed. (c) Marks of a sidewinder ventral scales during sidewinding, indicating static (peeling/rolling) contact.   }
     \label{FigS2}
 \end{suppfig}

  \newpage
 
     \begin{suppfig}[H]
    \begin{center}
            \includegraphics[width = 6.5in,keepaspectratio=true]
         {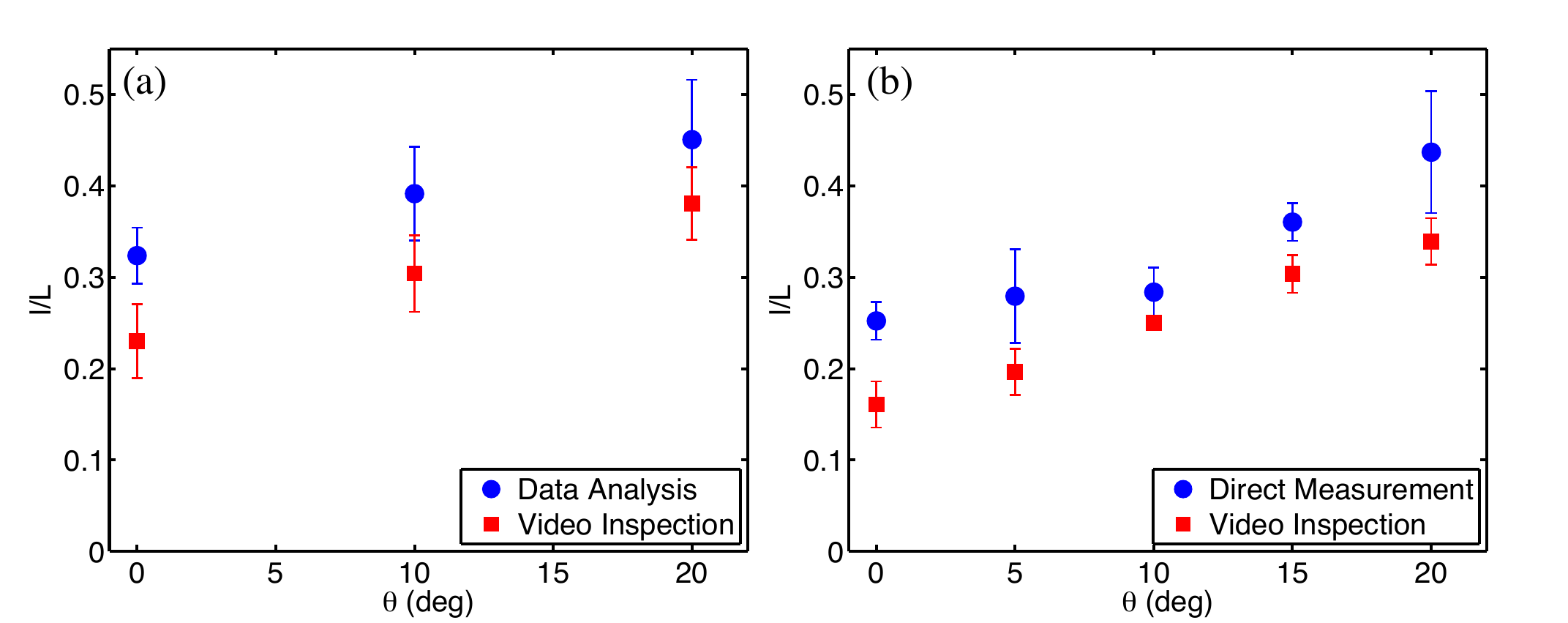}
    \end{center}
 \caption{ Comparison of data analyzed using the two different methods described in the supporting methods text for (a) snakes and (b) robot. Data denote mean +/- SD. }
     \label{FigS3}
 \end{suppfig}
 
  \newpage
 
     \begin{suppfig}[H]
    \begin{center}
            \includegraphics[width = 6.5in,keepaspectratio=true]
         {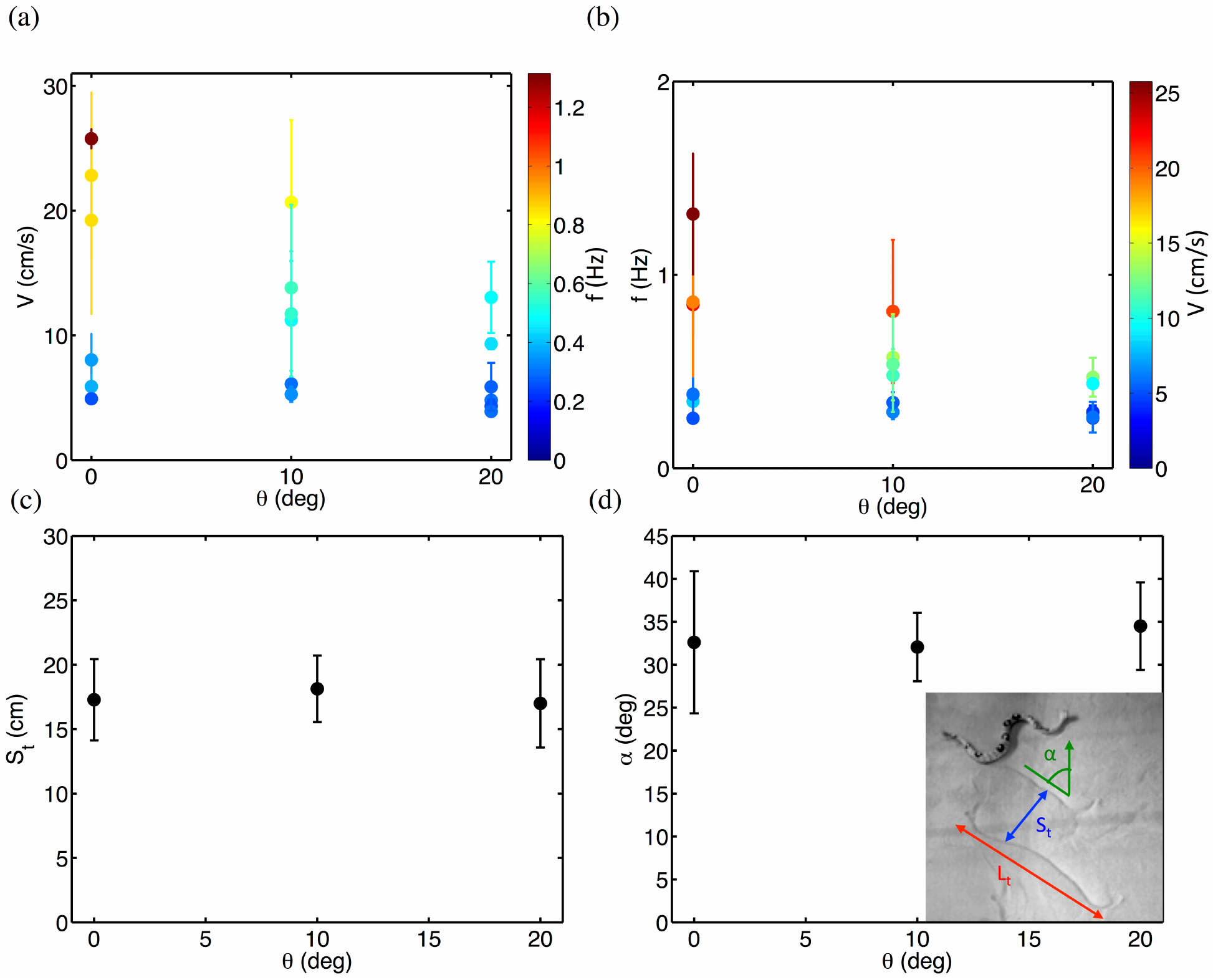}
    \end{center}
 \caption{Snake (a) speed $V$ color coded with wave frequency $f$, (b) wave frequency color coded with speed, (c) spacing of tracks $S_t$, and (d) angle between tracks and direction of motion $\alpha$ versus inclination angle $\theta$. Data denote mean $\pm$ SD.}
     \label{FigS4}
 \end{suppfig}
 
  \newpage
 
    \begin{suppfig}[H]
    \begin{center}
            \includegraphics[width = 6.5in,keepaspectratio=true]
         {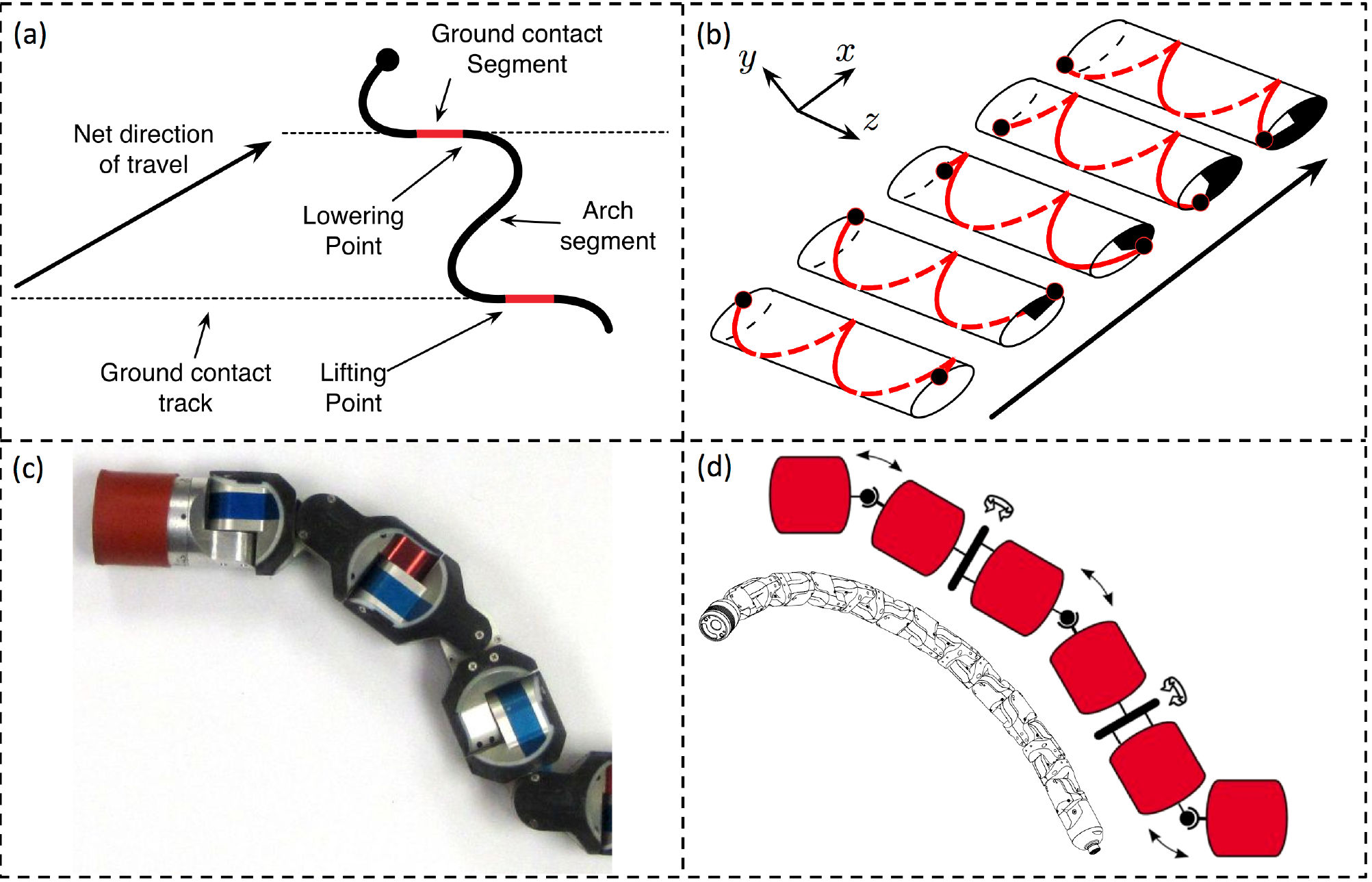}
    \end{center}
 \caption{(a) A schematic of the sidewinding gait showing the direction of travel, direction of traveling wave, and regions of locomotor in contact with ground \cite{gong2012conical}. (b) The sidewinding backbone curve is an elliptical helix which acts as a helical tread around a hypothetical core cylinder, driving it forward \cite{gong2012conical}. (c)-(d) The CMU robot. Each module has one degree of freedom and is oriented 90 degrees with respect to the previous one to allow motion in three dimensions. The kinematic configuration of the robot has single-DOF joints that are alternately oriented in the lateral and dorsal planes of the robot, with each joint having a full $\pm 90^{\circ}$ range of motion. The front module of the robot has a camera with LED lighting, and the tail module includes a tether connection with integrated slip-ring. }
     \label{FigS5}
 \end{suppfig}
 
  \newpage
 
     \begin{suppfig}[H]
    \begin{center}
            \includegraphics[width = 6.5in,keepaspectratio=true]
         {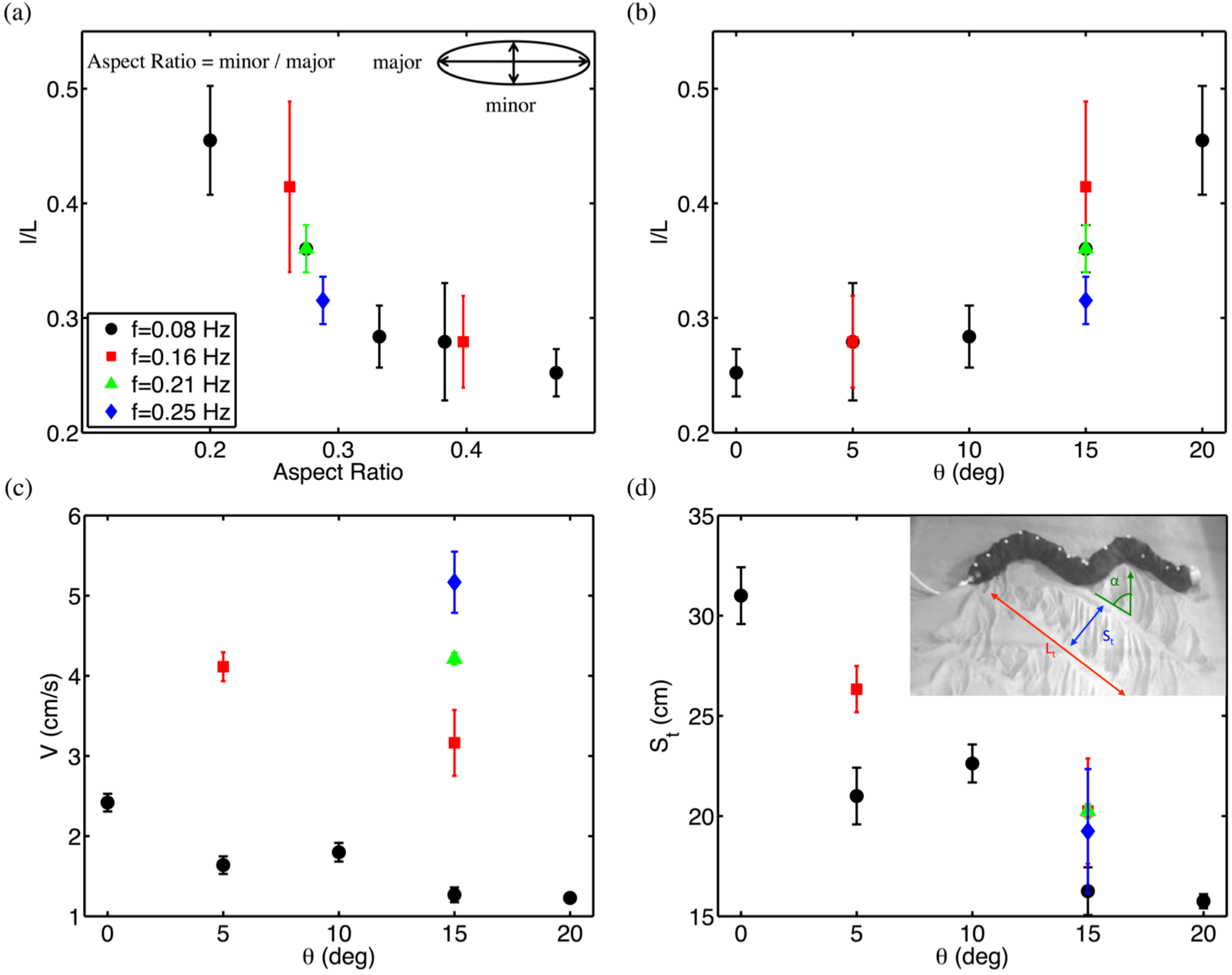}
    \end{center}
 \caption{(a) The normalized contact length $l/L$ of CMU snake robot versus aspect ratio of the elliptical helix at different frequencies. Snake robot's (b) normalized contact length $l/L$, (c) speed $V$, and (d) spacing of tracks $S_t$ versus inclination angle $\theta$. The data presented in (a)-(d) are for successful climbing trials (without slipping or pitching down the hill). Data denote mean $\pm$ SD. }
     \label{FigS6}
 \end{suppfig}
 
 \newpage
 
%      \begin{suppfig}[H]
%    \begin{center}
%            \includegraphics[width = 4in,keepaspectratio=true]
%         {Figures/FigS7.pdf}
%    \end{center}
% \caption{Experimental data compared to model predictions for pitching contact length versus $\theta$. Data are mean +/- SD.}
%     \label{FigS7}
% \end{suppfig}
 
 \newpage

   \begin{supptable}[H]
    \begin{center}
     \caption{The locomotor behavior of thirteen species of pit vipers on sand at zero and ten degree inclines. } 
            \includegraphics[width = 6in,keepaspectratio=true]
         {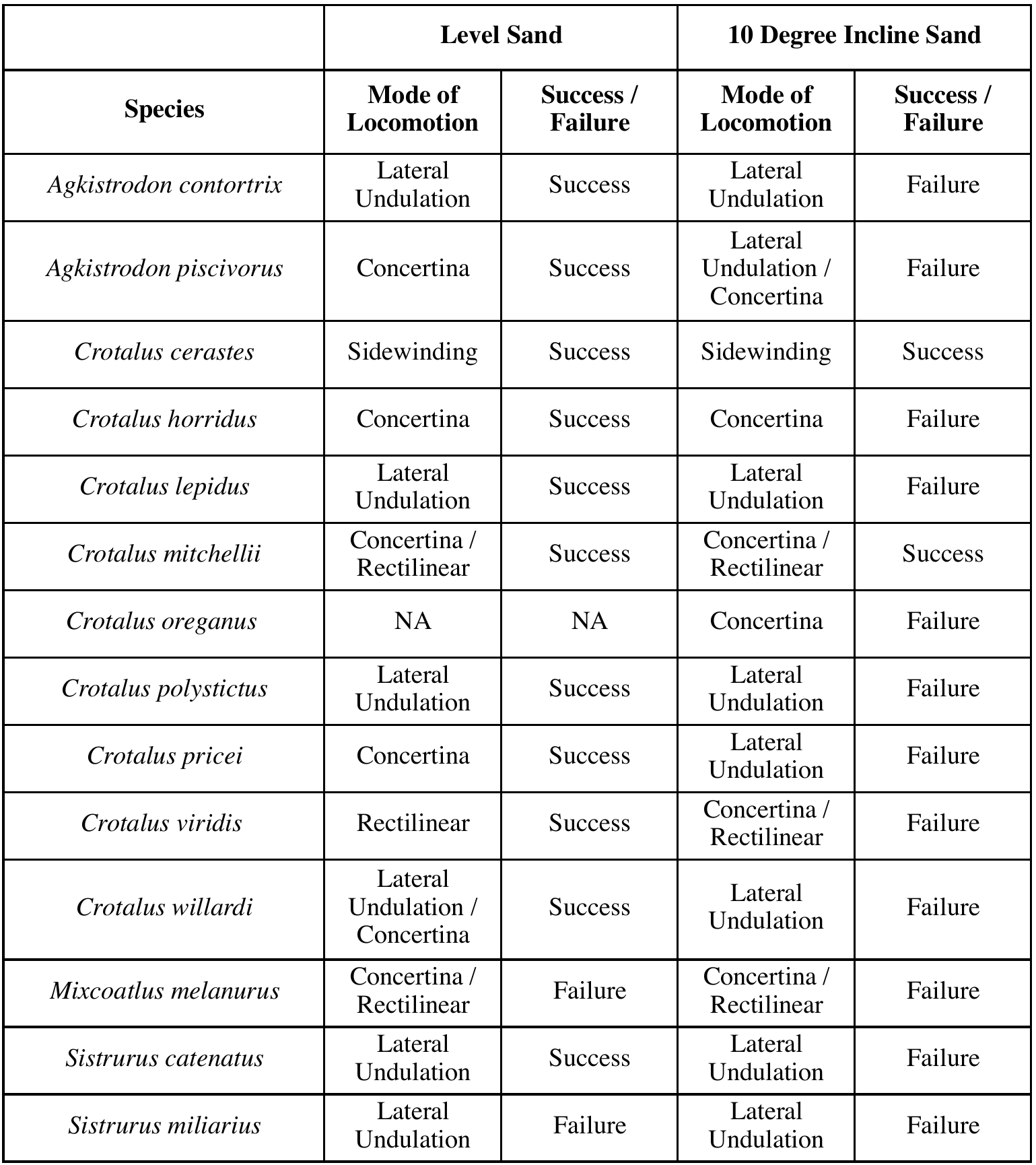}
    \end{center}

     \label{Table_S1}
 \end{supptable}

    \begin{supptable}[H]
    \begin{center}
     \caption{Particle size distribution of sand from Yuma County, Arizona \cite{li2013terradynamics}. } 
            \includegraphics[width = 4in,keepaspectratio=true]
         {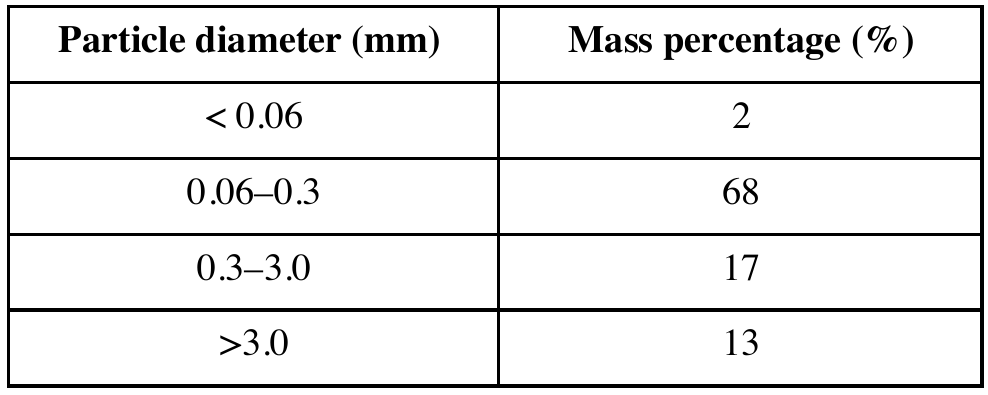}
    \end{center}

     \label{Table_S2}
 \end{supptable}

 \newpage

%%%%%%%%%%%%%%%%%%%%%%%%%%%%%%%%%%%%%%%%%%%%%%%%%%%%%%%%%%%
\section*{Supplementary Movies}

{\bf Movie S1} \\
\noindent 
A sidewinder rattlesnake climbing on loose sand. The videos illustrate sidewinding motion of a snake on inclinations of 0, 10, and 20$^\circ$ at real-time speed followed by 4-times slower speed. The side by side videos show each trial from two different angles. 

\noindent
{\bf Movie S2} \\
\noindent
The CMU robot climbing on loose sand at wave frequency of $f$=0.08 Hz. At inclination of 10 degrees the robot pitched at contact length of $l/L=0.28$ and slipped at $l/L=1$. CMU robot could successfully climb inclination of 10 degrees at $l/L=0.55$. The last two videos show the robot climbing inclinations of 5 and 20$^\circ$ at similar contact lengths successfully ($l/L=0.49$ and 0.45, respectively). However, due to the presence of local slipping at higher inclination angle  the step length is shorter and thus speed is slower ($\theta=20^\circ$).

\noindent
{\bf Movie S3} \\
\noindent
Movements of Other Crotaline Vipers on Horizontal and Inclined Sand. Sequence 1- A banded rock rattlesnake ({\it Crotalus lepidus}) uses lateral undulation on level sand. Sequence 2 - A speckled rattlesnake ({\it Crotalus mitchellii}) uses concertina with rectilinear locomotion on level sand (2x speed). Sequence 3 - A Mexican pitviper ({\it Mixcoatlus melanurus}) attempts to move using concertina locomotion on horizontal sand, but fails to make forward progress (3x speed).  Sequence 4 - A pigmy rattlesnake ({\it Sistrurus miliarius}) attempts to move using lateral undulation on horizontal sand, but fails to make forward progress. Sequence 5 - A speckled rattlesnake ({\it Crotalus mitchellii}) uses concertina with rectilinear locomotion on 10$^\circ$ inclined sand (2x speed).  Sequence 6 - A ridge-nosed rattlesnake ({\it Crotalus willardi}) attempts to move using lateral undulation on 10$^\circ$ inclined sand (uphill is upwards in the video), but fails to make forward progress. Sequence 7 - A Mexican pitviper ({\it Mixcoatlus melanurus}) attempts to move using concertina locomotion on 10$^\circ$ inclined sand, but fails to make forward progress.  Sequence 8 - A pigmy rattlesnake ({\it Sistrurus miliarius}) attempts to move using lateral undulation on 10$^\circ$ inclined sand (uphill is upwards in the video), but fails to make forward progress.

\noindent
{\bf Movie S4} \\
\noindent
Fluidizing sand using an air-fluidized bed. We constructed a setup to prepare a uniform and consistent state for the granular media before each trial. Our fluidized bed has a porous floor allowing the air to uniformly flow through the entire sand and letting it re-settle into an equilibrium condition. As shown in this video, regardless of the initial state of the sand we were able to achieve a loosely packed granular media with a smooth surface after the fluidization process.

\noindent
{\bf Movie S5} \\
\noindent
A sidewinder rattlesnake climbing loose sand at an inclination of 27$^\circ$. The video is sped up 4 times and illustrates the extended contacts the snake makes during sidewinding motion on the highest possible angle (angle of maximum stability) on loose sand.

\end{document}